\documentclass[journal]{IEEEtran}

\usepackage{amsmath,amsfonts}
\usepackage{algorithmic}
\usepackage{algorithm}
\usepackage{array}
\usepackage[caption=false,font=footnotesize]{subfig}
\usepackage{textcomp}
\usepackage{stfloats}
\usepackage{url}
\usepackage{verbatim}
\usepackage{graphicx}
\usepackage{xcolor}
\usepackage{soul}
\usepackage{cite}
\usepackage{makecell}

\makeatletter
\renewcommand{\section}{\@startsection{section}{1}{\z@}%
  {0.4ex plus 0.1ex minus 0.1ex}%
  {0.2ex plus 0.1ex minus 0.1ex}%
  {\normalfont\normalsize\centering\scshape}}

\renewcommand{\subsection}{\@startsection{subsection}{2}{\z@}%
  {0.3ex plus 0.1ex minus 0.1ex}%
  {0.15ex plus 0.1ex minus 0.1ex}%
  {\normalfont\normalsize\itshape}}
\makeatother

\begin{document}

\title{Exploiting Spatial Modulation for Strong Phase Noise Mitigation in mmWave Massive MIMO}

\author{
Oshin Daoud, Haïfa Farès, Amor Nafkha,~\IEEEmembership{Member,~IEEE},\\
Yahia Medjahdi, Laurent Clavier,~\IEEEmembership{Member,~IEEE}
\thanks{O. Daoud, H. Farès, and A. Nafkha are with IETR - UMR CNRS 6164, CentraleSupélec, 35576 Cesson-Sévigné, France (emails: \{oshin.daoud, haifa.fares, amor.nafkha\}@centralesupelec.fr).}
\thanks{Y. Medjahdi and L. Clavier are with IMT Nord Europe, Institut Mines-Télécom, Centre for Digital Systems, F-59653 Villeneuve d’Ascq, France (emails: \{yahia.medjahdi, laurent.clavier\}@imt-nord-europe.fr).}
\thanks{L. Clavier is with IEMN - UMR CNRS 8520, University of Lille, France.}
}

\markboth{IEEE Transactions on Vehicular Technology}%
{Daoud \MakeLowercase{\textit{et al.}}: Exploiting Spatial Modulation for Strong Phase Noise Mitigation in mmWave Massive MIMO}

\IEEEpubid{0000--0000/00\$00.00~\copyright~2026 IEEE}

\maketitle

\begin{abstract}

This letter investigates phase noise (PN) mitigation in generalized receiver spatial modulation (GRSM) massive MIMO systems at mmWave under a common local oscillator (CLO). Under CLO, the received energy remains invariant relative to the no-PN scenario, enabling reliable energy-based spatial detection using the no-PN threshold. PN-sensitivity and geometry-based metrics are introduced to design compact, PN-resilient MQAM symbol pools with low detection complexity. PN robustness is further improved through an enhanced PN-aware GRSM-MQAM system that exploits spatial modulation (SM) to recover part of the MQAM bits and strategically maps spatial-pattern Hamming weights to reduce the effective PN impact. In addition, a practical single-stage PN estimation/compensation architecture is proposed, while a benchmark double-stage compensation is adopted to quantify the upper bound achievable via separate Tx/Rx PN mitigation. Results show that under PN, the overall BER is mainly dominated by MQAM symbol detection errors, especially for denser constellations, whereas spatial detection remains robust. The proposed single-stage compensation improves PN resilience, while the benchmark double-stage compensation approaches near PN-free performance.

\end{abstract}
\section{Introduction}
\IEEEPARstart{I}{n} 5G mmWave (FR2) and emerging 6G upper-mid-band (FR3) systems, SM can improve spectral and energy efficiency. Although conventional transmitter SM (TSM) improves spectral efficiency, its reliance on antenna deactivation reduces beamforming gain, limiting performance at high frequencies with strong propagation losses.  
Receiver SM (RSM) and its generalized form (GRSM) activate one or $N_a$ receive antennas while preserving full transmit-side beamforming~\cite{b1,b2}, making GRSM well suited for FR2/FR3 systems. In~\cite{b3}, energy-based spatial detection revealed a performance advantage of GRSM-16QAM over GRSM-4QAM. This gain stems from its use of fewer spatial components, which reduces dominant spatial errors. However, the analysis did not consider hardware impairments.
Among these, PN from local oscillators is critical at high frequencies, causing spectral spreading and random rotations that degrade performance~\cite{b4,b5}. Prior work has primarily focused on low-order modulations such as QPSK/DQPSK and TSM~\cite{b6,b7}. In contrast, FSIM-NET~\cite{b8} uses filter-indexed modulation and achieves robust performance under severe PN due to its lightweight design.
Furthermore, PN induces an irreducible error floor, motivating mitigation strategies such as receiver-side compensation~\cite{b9} or constellation optimization~\cite{b10}. However, the aforementioned approaches often ignore hardware complexity, require prior PN knowledge, and, 
above all, neglect the key aspect of exploiting the spatial dimension for PN mitigation.
In this work, we study a 28~GHz massive MIMO downlink system employing GRSM-MQAM~\cite{b3}. Spatial detection is performed via energy sensing using a predefined no-PN threshold, while  MQAM symbol detection relies on conventional Maximum Likelihood (ML) based on Euclidean distance, without requiring PN statistics at the receiver. PN is modeled as a multivariate Gaussian process under strong CLO conditions, capturing practical RF impairments.
To the best of our knowledge, this is the first systematic analysis of PN in GRSM-MQAM. The main contributions are summarized as follows:
\textbf{(i)} It is proven that, under CLO operation, the received energy remains identical to the no-PN case, ensuring that the predefined no-PN threshold remains valid for robust spatial detection even under strong PN.
\textbf{(ii)} MQAM symbols are newly classified into PN-robust and PN-sensitive groups. In addition, a geometric criterion is introduced to minimize PN-induced overlap by maintaining a $\pi$ angular separation within each low-complexity MQAM symbol pool.
\textbf{(iii)} An enhanced PN-aware scheme (E-PN-GRSM-MQAM) is proposed based on the proposed pools, exploiting SM as an additional degree of freedom for PN mitigation. 
\textbf{(iv)} Symbol-assisted PN estimation and compensation procedure is introduced for E-PN-GRSM-MQAM under two architectures: a practical single-stage and a benchmark double-stage scheme.
\IEEEpubidadjcol
\section{System Model}
\subsection{The Transmitter}
We consider a time-division duplex (TDD) massive MIMO (mMIMO) system where channel reciprocity enables the use of uplink channel estimates for downlink transmission, as in \cite{b3}. The system consists of a single-cell base station (BS),  serving as the transmitter,  with $N_t$ antennas and $N_t$ RF chains serving a single user with $N_r$ receive antennas.
In a classical (spatial-MQAM) mapping, the incoming binary stream is partitioned into two parts: the first $N_a$ bits represent the spatial pattern, the remaining $\log_2(M)$ bits encode the MQAM-modulated symbol. The resulting spectral efficiency is  
$\text{SE} = N_a + \log_2(M)$ \text{bits/sec/Hz}.
In this work, all illustrations are given for a fixed $\text{SE}=8$ \text{bits/sec/Hz}, the number of spatial bits is chosen as $N_a = $ 4 or 6, for $M =$ 16 or 4, respectively.
Assuming perfect channel knowledge at the BS, receiver antenna selection algorithm is employed to choose the $N_a$ least correlated receive antennas from the set of $N_r$ available antennas.
As a result, the channel matrix $\mathbf{H} \in \mathbb{C}^{N_r \times N_t} $ is reduced to $ \mathbf{H}_a \in \mathbb{C}^{N_a \times N_t}$.
 The $N_a$ bits of each spatial pattern are then mapped to the selected receive antennas using a Zero-Forcing precoder $\mathbf{B} \in \mathbb{C}^{N_t \times N_a}$. 
The transmitter PN effect is given by the vector
$\boldsymbol{\Phi}_{\text{tx}} = \operatorname{diag}\left( \left[e^{j\phi_1}, \cdots, e^{j\phi_{N_t}} \right]^{\mathrm{T}} \right).$
The channel follows an extended Saleh-Valenzuela model to capture 28~GHz mmWave propagation as a narrowband clustered channel with line-of-sight and no-line-of-sight conditions \cite{b3}.
\subsection{The Receiver}
The received signal at the $N_a$ active receiver branches:
\begin{equation}
    \mathbf{y} = \sqrt{\alpha } \,  \mathbf{H}_a\boldsymbol{\Phi}_{\text{tx}} \mathbf{B} s_i x_j + \mathbf{n},
    \label{eq:1}
\end{equation}
where $\mathbf{y} \in \mathbb{C}^{N_a \times 1} $ denotes the received signal vector, and
$\mathbf{n} \in \mathbb{C}^{N_a \times 1} \sim \mathcal{CN}\!\left(\mathbf{0}, \sigma^2 \mathbf{I}_{N_a}\right)$ the complex additive white Gaussian noise (AWGN) vector, with zero-mean and $\sigma^2 \mathbf{I}_{N_a}$ is the covariance matrix, with $\mathbf{I}_{N_a}$ denoting the $N_a \times N_a$ identity matrix.
The average power normalization factor is given by
$\alpha
= \mathbb{E}_{\mathbf H_a}
\!\left\{
\big[\operatorname{tr}\!\big((\mathbf H_a \mathbf H_a^{H})^{-1}\big)\big]^{-1}
\right\}.$
The modulated symbol used in MQAM modulation belongs to the constellation
$\mathcal{C} = \{ x_j \mid j \in \{1, .., M\} \},$ each has the phase $\theta_x$ and the amplitude $|x_j|$, represented in $m = \log_2(M)$ bits,
and the spatial pattern is defined as
$\mathbf{s}_i \in \{0,1\}^{N_a \times 1},$
where $ i \in \{1,.., 2^{N_a} - 1\}$, assuming the all-zeros pattern is excluded to avoid deactivating all selected receive antennas.
Given $\mathbf{s}_i$,  we define its Hamming weight as
\(
w_{h,i} = \sum_{k=1}^{N_a} s_{ik}
\), where $s_{ik}$ the $k_{th}$ bit of $\mathbf{s}_i$.
Thus, $w_h\in\{1,..,N_a\}$.
For ease of tracking, we express each $\mathbf{s}_i$ using its decimal value $J\in\{1,..,2^{N_a}-1\}$.
In the case of a CLO, PN effect is
$\boldsymbol{\Phi}_{\text{tx}} =  e^{j\phi} \mathbf{I}_{N_t},$ thus the received signal  on the $k_{th}$ branch:
\begin{equation}
y_k = \sqrt{\alpha}\, |x_j| \, e^{j \theta_x} e^{j \phi}\, s_{ik} + n_k.
\label{eq:2}
\end{equation}
The detector first recovers the spatial pattern $\mathbf{s}_i$, with each bit $s_{ik}$ obtained from its corresponding received branch $y_k$. Specifically, an Energy Detector measures the PN-affected energy $z_k$, and a 1-bit ADC compares it to a threshold $\gamma$: $\hat{s}_{ik} = 0$ if $z_k < \gamma$, $\hat{s}_{ik} = 1$ otherwise.
The energy detection is implemented in the RF domain to enable a low-complexity and energy-efficient receiver architecture \cite{b11}. Nevertheless, we perform the threshold design and related derivations in the equivalent baseband domain. This is justified because the energy of the signal is preserved under frequency translation (Parseval theorem), and the statistical distributions of $z_k$ under the hypotheses $s_{ik} \in \{0,1\}$ are unchanged in the baseband representation as shown in \cite{b12}.
Second, once the spatial pattern has been recovered, the detection of the MQAM symbols is carried out by combining the $N_a$ received active branches, followed by a conventional  ML detector,
using:
\begin{equation}
\hat{x}_j = \arg\min_{x_j \in \mathcal{C}} |y_c - x_j|^2,
\label{eq:3}
\end{equation}
where
$y_c = e^{j\phi_{\mathrm{rx}}} \sum_{k=1}^{N_a} \hat{s}_{ik} y_k$
is the combined signal and, passing through a single RF chain, it experiences one sample of PN, denoted by $\Phi_{\mathrm{rx}} = e^{\,j \phi_{\mathrm{rx}}}$.
\section{Phase Noise Modelling}
Under the CLO configuration, the PN samples across all transmit RF chains are fully correlated.
At mmWave and sub-THz frequencies, the nearly flat PN power spectral density over the signal bandwidth implies negligible temporal correlation in high-rate communications, \cite{b4,b5,b6}.
Accordingly, the PN samples are accurately modeled  as a multivariate Gaussian random process with distribution $\mathcal{N}(\mathbf{0} , \boldsymbol{\Sigma})$\cite{b4},
where the covariance matrix for our fully digital BS ($N_t$ RF chains):
\begin {equation}
\boldsymbol{\Sigma} =
{\small
\begin{bmatrix}
\sigma_{\mathrm{PN},1}^2 & \rho & \cdots & \rho \\
\vdots & \ddots & \ddots & \vdots \\
\rho & \cdots & \rho & \sigma_{\mathrm{PN},N_t}^2
\end{bmatrix}
}
\label{eq:4}
\end {equation}
Here, $\rho$ denotes the correlation coefficient between PN samples of different RF chains. For the CLO, the correlation coefficient is set to $\rho = 1$, corresponding to fully correlated PN.
At the receiver, a single PN sample is generated and modeled as a zero-mean Gaussian random variable with variance $\sigma_{\mathrm{PN}}^2$.
\section{The received signal model with PN}
In \cite{b3}, the received signal at each antenna is analyzed in terms of its real and imaginary components for both spatial bit hypotheses, $s_{ik} \in \{0,1\}$. In the absence of PN, $y_k \mid s_{ik} \sim \mathcal{CN}(\mu_k,\sigma^2)$, where $\mu_k = 0$ for $s_{ik}=0$ and $\mu_k = \sqrt{\alpha}\,x_j$ for $s_{ik}=1$. With PN, from \eqref{eq:2}, when $s_{ik}=0$, $y_k=n_k$ and the normalized received energy 
$z_k = 2|y_k|^2/\sigma^2$ follows a central chi-squared distribution with 2 degrees of freedom.
When $s_{ik} = 1$, we show that the received energy can still be approximated as a noncentral chi-square random variable.
Indeed, based on \eqref{eq:2}, letting $A = \sqrt{\alpha}|x_j|$,
and using a first-order Taylor series approximation under the assumption that $\sigma_\text{PN}^2 \ll 1$ to ignore higher-order terms, the real and imaginary parts of the received signal are given by
\begin{equation}
\left\{
\begin{aligned}
y_{k,\mathrm{re}} \approx A\!\left(\cos\theta_x - \phi\sin\theta_x\right) + n_{k,\mathrm{re}},\\
y_{k,\mathrm{im}} \approx A\!\left(\sin\theta_x + \phi\cos\theta_x\right) + n_{k,\mathrm{im}}.
\end{aligned}
\right.
\label{eq:5}
\end{equation}
Thus, $y_{k,\mathrm{re}}$ and $y_{k,\mathrm{im}}$ are linear transformation of the zero-mean Gaussian random variable $\phi$, preserving the Gaussianity. 
Consequently, the received energy is noncentral chi-square distributed. 
Under a CLO configuration, PN induces a common phase rotation across all $N_t$ transmit branches, which does not affect the signal magnitude at each receive antenna. Consequently, the received energy remains unchanged, explaining its equivalence to the no-PN case. 
\section{PN-aware MQAM symbols classification}
\subsection{Taylor series-based classification}
Each transmitted MQAM symbol $x_j$ under PN can be expressed as
$ x_j^{\rm PN} = |x_j| e^{j(\theta_x + \phi)},$
where $\phi$ is the PN perturbation. 
Its real and imaginary components can be approximated via a first-order Taylor expansion:
$ x_j^{\rm PN} \approx |x_j|(\cos\theta_x - \phi\sin\theta_x) + j|x_j|(\sin\theta_x + \phi\cos\theta_x).$ In 16QAM, diagonal symmetric symbols ($\pm 45^\circ, \pm 135^\circ$) have equal-magnitude sine and cosine values, so PN perturbs their real and imaginary parts evenly. Off-diagonal asymmetric symbols lack this symmetry, resulting in unbalanced deviations. 
Relative amount of perturbation introduced by the PN effect on the real and imaginary components $\Re\{\cdot\}$ and $\Im\{\cdot\}$, denoted
$\varepsilon_{\rm re}$, $\varepsilon_{\rm im}$, respectively:
\begin{equation}
\varepsilon_{\{\rm re,im\}}(x_j)[\%]=
\tfrac{\{\Re,\Im\}\!\left\{x_j^{\rm PN}\right\}-\{\Re,\Im\}\!\left\{x_j^{\rm no-PN}\right\}}
{\{\Re,\Im\}\!\left\{x_j^{\rm no-PN}\right\}} \times 100.
\label{eq:6}
\end{equation}
At $\phi = 0.1$, $x_j = 3-3i$ experiences
$\varepsilon_{\rm re} = \varepsilon_{\rm im} = 10\%$, while $x_j = -3-1i$ shows $\varepsilon_{\rm re} \approx 3.33\%, \varepsilon_{\rm im} = 30\%$.  
Based on this, 16QAM symbols are classified into a PN-robust diagonal group $\mathcal{R}$ and a PN-sensitive off-diagonal group $\mathcal{S}$. In contrast, 4QAM symbols are all symmetric and equally robust.
\subsection{Geometrical classification (PN overlap minimization)}
In MQAM systems operating under PN–dominated conditions, detection performance is primarily governed by the angular separation between constellation points rather than their Euclidean distance. Let two transmitted symbols be characterized by phases $\theta_{x_1}$ and $\theta_{x_2}$. Assuming a common PN impairment modeled as a random phase rotation $\phi \sim \mathcal{N}(0,\sigma_{\rm PN}^2)$, the observed symbol phases follow Gaussian distributions with means $\theta_{x_1}$ and $\theta_{x_2}$ and identical variance $\sigma_{\rm PN}^2$.
The overlap probability is defined as a measure of the statistical overlapping between the two resulting phase distributions. Given independent PN realisations affecting $\theta_{x_1}$ and $\theta_{x_2}$, the overlap probability is given by the integral of the product of their probability density functions (PDFs):
\begin{equation}
P_{\text{Overlap}}
= \int_{-\infty}^{\infty} f_1(\phi)\,f_2(\phi)\,d\phi,
\label{eq:7}
\end{equation}
where $f_i(\phi)=\mathcal{N}(\theta_{x_i},\sigma_{\rm PN}^2)$ for $i=1,2$. This quantity provides an analytical indicator of the likelihood that PN causes ambiguity between the two symbols.
Evaluating the integral yields the following closed-form expression:
\begin{equation}
P_{\text{Overlap}}
= \frac{1}{2\sqrt{\pi}\,\sigma_{\rm PN}}
\exp\!\left(-\frac{\Delta\theta^2}{4\sigma_{\rm PN}^2}\right),
\label{eq:8}
\end{equation}
where $\Delta\theta = |\theta_{x_1}-\theta_{x_2}|$ denotes the angular separation between the symbols. This expression highlights the exponential decay of the overlap probability with increasing angular separation.
For a 4QAM constellation, symbols located in opposite diagonal quadrants achieve the maximum angular separation $\Delta\theta=\pi$. As a numerical illustration, for a strong PN variance $\sigma_{\rm PN}^2=0.1~\text{rad}$, the overlap probability evaluates to
\(
P_{\text{Overlap}}
\approx \frac{1}{2\sqrt{\pi}\,\sigma_{\rm PN}}
\exp\!\left(-\frac{\pi^2}{4\sigma_{\rm PN}^2}\right)
\approx 5.4 \times 10^{-11},
\)
which is negligibly small. Based on this result, a reduced-complexity structure (pool) is proposed, where symbols separated by $\pi$ are paired to minimize the probability of PN-induced overlap.
\subsection{Proposed Pools Construction}
\begin{figure}[!t]
\centering
\subfloat[]{%
    \includegraphics[width=0.5\columnwidth]{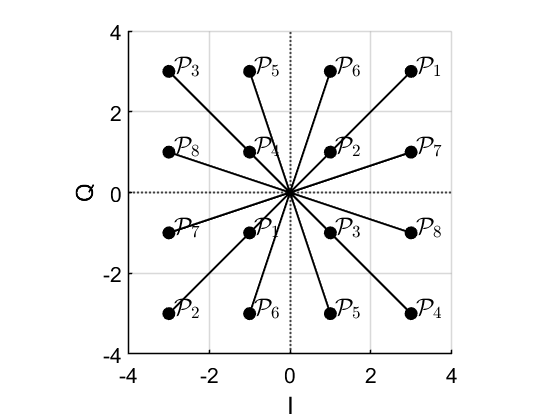}%
    \label{fig:1a}%
}
\hfill
\subfloat[]{%
    \includegraphics[width=0.5\columnwidth]{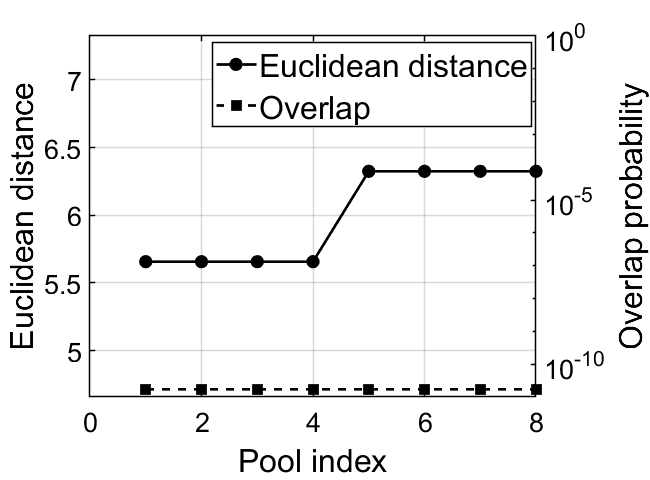}%
    \label{fig:1b}%
}
\caption{(a) 16QAM symbol pools construction.
(b) Euclidean distance and overlapping probability for 16QAM pools.}
\label{fig:1}
\end{figure}

Based on both classification criteria, a pooling strategy is proposed for an MQAM constellation ($M=2^m$). The constellation is divided into $M/2=2^{m-1}$ disjoint pools, indexed from $0$ to $2^{m-1}-1$, with each pool containing two symbols. This paired-symbol structure reduces the search space.
To minimize the overlap probability, symbol pairing prioritizes the largest angular separation, ideally $\pi$, by selecting diagonally opposite constellation points. The Euclidean distance is used as a secondary criterion and is maximized under this angular-separation constraint.
For 4QAM, all symbols have the same PN sensitivity. Therefore, only the geometric criterion is needed, yielding two pools, $\mathcal{P}_1$ and $\mathcal{P}_2$, each formed by a pair of symbols with maximum angular separation, as listed in Table~\ref{tab:4QAM_mapping}.
For 16QAM, symbols exhibit different PN-sensitivity levels. Hence, the pool construction must satisfy both criteria: (1) each pool contains two diagonally opposite symbols to reduce overlap probability ($\pi$ separation) (2) both symbols in the same pool belong to the same PN-sensitivity class. Accordingly, the constellation is divided into 8 pools. Four pools containing diagonal PN-robust symbols, $\mathcal{P}_1$--$\mathcal{P}_4$, are grouped into $\mathcal{R}$, with pairs defined as $\{x_r,\,x_{\frac{\sqrt{M}}{2}+r}\}$:  $r=1,..,\frac{\sqrt{M}}{2}$. The other four pools, $\mathcal{P}_5$--$\mathcal{P}_8$, contain PN-sensitive symbols and are grouped into $\mathcal{S}$, where the two symbols in each pool are antipodal (symmetric about the origin).
The resulting pools are summarized in Table~\ref{tab:16QAM_mapping} and illustrated in Fig.~\ref{fig:1a}. Under these rules, the overlap probability is minimized through diagonally opposite pairing, while the Euclidean distance is optimized within each pool, as in Fig.~\ref{fig:2a}.

\section{Enhanced PN-aware GRSM-MQAM System (E-PN-GRSM-MQAM)}
The proposed E-PN-GRSM-MQAM system exploits SM as an additional degree of freedom to mitigate PN in two ways: (i) it leverages the inherent PN robustness of spatial detection to reliably recover part of the MQAM bits, (ii) it uses the Hamming weights of spatial patterns in the pool-to-pattern mapping to reduce the effective PN impact.
Unlike classical mapping, where MQAM symbols and spatial patterns are treated as independent streams, the proposed scheme adopts an MQAM-symbol-driven spatial-pattern mapping strategy. Specifically, the MQAM bit sequence $\mathbf{b}=\big(b_1,\dots,b_m\big)$ is divided such that $b_1,\dots,b_{m-1}$ select the pool index, while $b_m$ selects the symbol within the chosen pool $\mathcal{P}$.
Once the pool $\mathcal{P}$ is selected, a binary spatial pattern $\mathbf{s}_i$, represented by its decimal index $J$, is assigned to that pool according to one constraint only: the Hamming weight $w_h$ of its binary representation.
For E-PN-GRSM-4QAM, $J$ takes values from $1$ to $63$. For $\mathcal{P}_1$, any index can be assigned provided that its binary representation has a low Hamming weight, $w_h\in\{1,2,3\}$. For $\mathcal{P}_2$, any index can be assigned provided that it has a high Hamming weight, $w_h\in\{4,5,6\}$. Since both pools have the same PN sensitivity, these Hamming-weight assignments can be interchanged without performance loss. The mapping is in Table~\ref{tab:4QAM_mapping}.
As an example, consider the 4QAM input bits \texttt{10}. The first bit, $b_1=1$, selects pool $\mathcal{P}_2$. Then, an index $J$ is randomly chosen such that the corresponding binary spatial pattern satisfies $w_h\in\{4,5,6\}$. Finally, the second bit, $b_2=0$, selects the first symbol in $\mathcal{P}_2$, yielding the transmitted symbol $\{1+1i\}$.
Unlike 4QAM, 16QAM exhibits non-uniform PN sensitivity; therefore, the Hamming-weight assignment must follow the PN classification of the pools. In E-PN-GRSM-16QAM, the decimal index $J$ ranges from $1$ to $15$. For the PN-robust pools $\mathcal{P}_1$--$\mathcal{P}_4$, any index may be assigned provided that its binary representation has a low Hamming weight, $w_h\in\{1,2\}$. For the PN-sensitive pools $\mathcal{P}_5$--$\mathcal{P}_8$, any index may be assigned provided that it has a high Hamming weight, $w_h\in\{3,4\}$. The mapping is in Table~\ref{tab:16QAM_mapping}.
For example, consider the input bits \texttt{1001}. The first three bits, $b_1b_2b_3=\texttt{100}$, select pool $\mathcal{P}_5\in\mathcal{S}$. Then, an index $J$ is randomly chosen from $\{7,13\}$ such that the corresponding spatial pattern satisfies $w_h=3$. Finally, $b_4=1$ selects the second symbol in $\mathcal{P}_5$, yielding the transmitted symbol $\{1-3i\}$.
At the receiver, the estimated spatial pattern $\mathbf{\hat{s}}_i$ yields the estimated decimal index $\hat{J}$, which determines the estimated pool $\hat{\mathcal{P}}$ from the mapping tables. Because the pool index encodes the first $(m-1)$ bits of the MQAM sequence, this step directly recovers them. A low-complexity ML search is then performed within the estimated pool to detect the transmitted symbol, whose local index determines the $m$th bit of the MQAM bits.
Finally, the assignment of high-Hamming-weight spatial patterns to the sensitive pools is motivated by the fact that activating more receive branches increases the diversity gain, which reduces the effective impact of PN. Furthermore, owing to the robustness of spatial detection under PN, a transmitted spatial pattern $\mathbf{s}_i$ with high Hamming weight is likely to be detected as a pattern $\hat{\mathbf{s}}_i$ that retains the same property. On this basis, the reduced PN impact can be proved analytically as follows:
For $N_a$ active receiver branches, the combined PN term is:
\begin{equation}
   R_{\rm PN}=e^{j\phi_{\rm rx}}\frac{\sum_{k=1}^{N_a} e^{j\phi_k}\hat{s}_{ik}}{\max\{\sum_{k=1}^{N_a}\hat{s}_{ik},1\}},
   \label{eq:9}
\end{equation}
where $\phi_{\rm rx}$ and $\phi_k$ denote the Rx-PN and the Tx-PN affecting the $k$th active branch, respectively, and are independent with distribution $\mathcal{N}(0,\sigma_{\rm PN}^2)$.
For example, with one active branch (pattern \texttt{1000}), $R_{\rm PN}=e^{j\phi_{\rm rx}}e^{j\phi_1}$, and under a first-order Taylor approximation, the effective phase perturbation is $\Theta_{R_{\rm PN}}\approx\phi_{\rm rx}+\phi_1$, giving $\mathrm{Var}(\Theta_{R_{\rm PN}})=2\sigma_{\rm PN}^2$. For two, three, and four active branches, the variance becomes $\frac{3}{2}\sigma_{\rm PN}^2$, $\frac{4}{3}\sigma_{\rm PN}^2$, and $\frac{5}{4}\sigma_{\rm PN}^2$, respectively. Since these values decrease monotonically with the number of active branches, higher Hamming-weight spatial patterns are more resilient to PN due to the increased diversity gain.

\begin{table}[!t]
\caption{E-PN-GRSM-4QAM mapping table.}
\label{tab:4QAM_mapping}
\vspace{1pt}
\centering
\footnotesize
\setlength{\tabcolsep}{3pt}
\renewcommand{\arraystretch}{1.1}
\begin{tabular}{|c|c|c|c|}
\hline
$b_1$ & $\mathcal{P}$ & Symbols & $J_{\text{Decimal}} = (\mathbf{s}_i)_{\text{Binary}}$ \\
\hline
0 & $\mathcal{P}_1$ & $[-1+1i,\,1-1i]$ &
$\{\,J \in \{1,\ldots,63\}\mid w_h(\mathbf{s}_i)=1,2,3\,\}$ \\
\hline
1 & $\mathcal{P}_2$ & $[1+1i,\,-1-1i]$ &
$\{\,J \in \{1,\ldots,63\}\mid w_h(\mathbf{s}_i)=4,5,6\,\}$ \\
\hline
\end{tabular}
\end{table}

\begin{table}[!t]
\caption{E-PN-GRSM-16QAM mapping table.}
\label{tab:16QAM_mapping}
\vspace{1pt}
\centering
\footnotesize
\setlength{\tabcolsep}{3pt}
\renewcommand{\arraystretch}{1.05}
\begin{tabular}{|c|c|c|p{3.75cm}|}
\hline
$b_1 b_2 b_3$ & $\mathcal{P}$ & Symbols & $J_{\text{Decimal}} = (\mathbf{s}_i)_{\text{Binary}}$ \\
\hline
000 & $\mathcal{P}_1$ ($\mathcal{R}$) & $[3+3i,\,-1-1i]$ 
& $\{\,J \in \{1,2\}\mid w_h(\mathbf{s}_i)=1\,\}$ \\
\hline
001 & $\mathcal{P}_2$ ($\mathcal{R}$) & $[-3-3i,\,1+1i]$ 
& $\{\,J \in \{4,8\}\mid w_h(\mathbf{s}_i)=1\,\}$ \\
\hline
010 & $\mathcal{P}_3$ ($\mathcal{R}$) & $[-3+3i,\,1-1i]$ 
& $\{\,J \in \{3,5,6\}\mid w_h(\mathbf{s}_i)=2\,\}$ \\
\hline
011 & $\mathcal{P}_4$ ($\mathcal{R}$) & $[-1+1i,3-3i]$ 
& $\{J\!\in\!\{9,10,12\}\mid w_h(\mathbf{s}_i)=2\}$ \\
\hline
100 & $\mathcal{P}_5$ ($\mathcal{S}$) & $[-1+3i,\,1-3i]$ 
& $\{\,J \in \{7,13\}\mid w_h(\mathbf{s}_i)=3\,\}$ \\
\hline
101 & $\mathcal{P}_6$ ($\mathcal{S}$) & $[1+3i,\,-1-3i]$ 
& $\{\,J = 11\mid w_h(\mathbf{s}_i)=3\,\}$ \\
\hline
110 & $\mathcal{P}_7$ ($\mathcal{S}$) & $[3+1i,\,-3-1i]$ 
& $\{\,J = 14\mid w_h(\mathbf{s}_i)=3\,\}$ \\
\hline
111 & $\mathcal{P}_8$ ($\mathcal{S}$) & $[-3+1i,\,3-1i]$ 
& $\{\,J = 15\mid w_h(\mathbf{s}_i)=4\,\}$ \\
\hline
\end{tabular}
\end{table}

\section{Symbol-Assisted PN Estimation and Compensation in E-PN-GRSM-MQAM}
\begin{figure}[!t]
\centering

\subfloat[]{%
    \includegraphics[width=0.9\columnwidth]{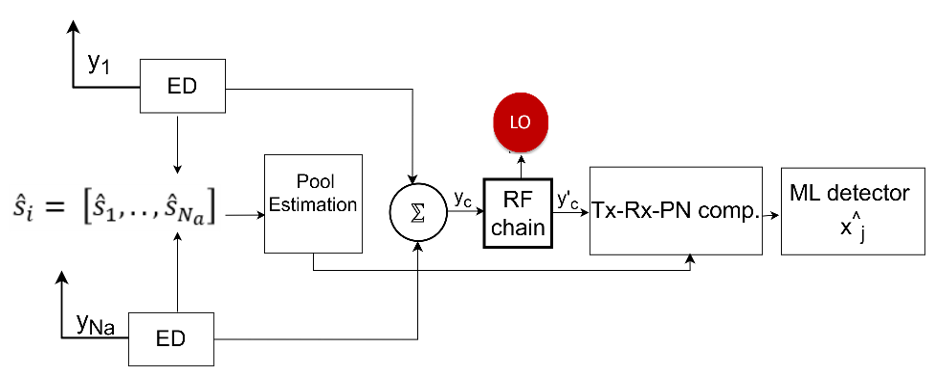}%
    \label{fig:2a}%
}
\vspace{-5mm} 
\subfloat[]{%
    \includegraphics[width=0.9\columnwidth]{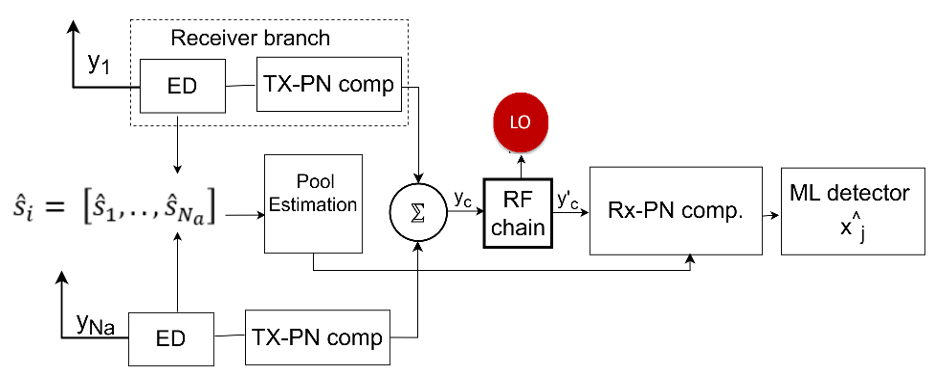}%
    \label{fig:2b}%
}
\caption{ Proposed structure: (a) practical single-stage compensation (b) benchmark double-stage compensation.}

\label{fig:2}
\end{figure}

This scheme exploits a key feature of E-PN-GRSM-MQAM: symbols are transmitted within known, compact pools with $\pi$ angular separation. This enables pilot-free PN estimation by measuring the angular mismatch between the received symbol and the closest candidate symbol in the estimated pool. 
Two architectures are considered. The single-stage (Fig.~\ref{fig:2a}), symbol-assisted PN estimation and compensation are applied once after spatial combining. The double-stage (Fig.~\ref{fig:2b}), they are applied twice: first per branch before combining to compensate analog-domain Tx-PN, and then after combining to compensate the common Rx-PN. The latter is used as a benchmark for separate Tx/Rx-PN mitigation.

\subsection{Single-Stage PN Estimation and Compensation}

After spatial combining, the received signal is affected by both Tx-PN (averaged over the active branches) and Rx-PN, which is common to all branches. These impairments appear as a single unknown phase rotation on the combined symbol. Owing to the E-PN-GRSM-MQAM design, detection is confined to a small known symbol pool whose elements are well separated in angle, reducing PN-induced ambiguity.
Formally, the combined signal is affected by the receiver oscillator PN as
$y_c' = y_c e^{j\phi_{\rm rx}}.$
A tentative symbol decision is then obtained by minimum-distance detection within $\mathcal{\hat{P}}$:
\begin{equation}
x_j' = \arg\min_{x_j \in \mathcal{\hat{P}}} |y_c' - x_j|^2.
\label{eq:10}
\end{equation}
This tentative decision serves as a phase reference. The PN-induced phase error is estimated as
$\Delta \phi = \angle y_c' - \angle x_j',$
and wrapped into the interval $(-\pi/2,\pi/2)$ as
$\Delta \phi_w = \mathrm{mod}(\Delta \phi + \pi/2, \pi) - \pi/2.$
This wrapping exploits the $\pi$ angular separation of the symbols within each pool, so that the correct decision region lies within half this separation. The estimated phase is then removed, and symbol detection is repeated in $\mathcal{\hat{P}}$:
\begin{equation}
\hat{x}_j' = \arg\min_{x_j \in \mathcal{\hat{P}}} \big| y_c' e^{-j\Delta \phi_w} - x_j \big|^2 ,
\label{eq:11}
\end{equation}
thereby improving robustness against moderate PN.
Under strong PN, the phase rotation may exceed $\pi/2$, making the wrapping ambiguous. In addition, joint Tx/Rx-PN estimation after combining limits mitigation. These limitations motivate the proposed double-stage structure.

\subsection{Double-Stage PN Estimation and Compensation}
Since Tx-PN affects each receive branch prior to spatial combining, the double-stage architecture first estimates and compensates this component on a per-branch basis using symbol-assisted processing:
\(
y_k e^{-j\Delta \phi_w}.
\)
Performing this operation before combining avoids masking severe phase deviations across branches. After this first stage, spatial combining is applied, and the resulting signal is then impaired only by the common Rx-PN. This remaining phase component can be estimated and compensated using the same angular-based symbol-assisted method. By treating the transmitter- and receiver-induced phase fluctuations separately, each estimation stage is more likely to operate within the interval $(-\pi/2,\pi/2)$, which improves estimation reliability.
This early-RF-domain processing is used as a benchmark, not as a straightforward hardware implementation, since it enables a clean separation of Tx-PN and Rx-PN before downconversion merges both effects. Moreover, \cite{b13} supports the feasibility of pre-baseband PN mitigation through analog electrical PN compensation.

\section{Implementation and Results}

\begin{figure}[!t]
\centering
\subfloat[]{%
        \includegraphics[width=0.9\columnwidth]{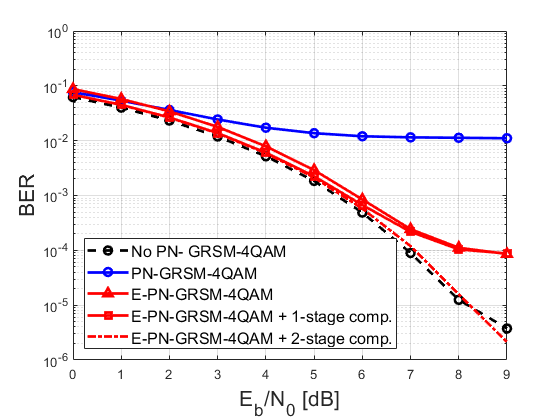}%
        \label{fig:3a}
    }
    \vspace{-1mm}
    \subfloat[]{%
        \includegraphics[width=0.9\columnwidth]{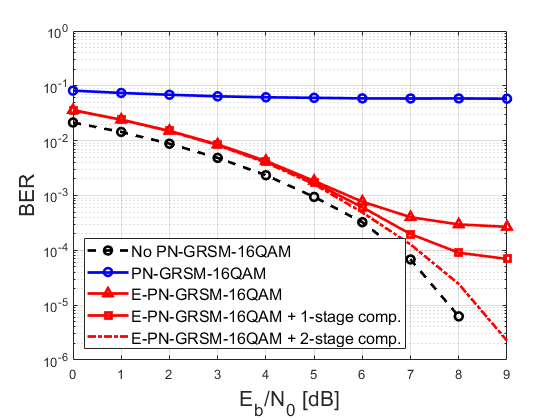}%
        \label{fig:3b}
    }
\caption{
 Overall BER for all system variants: 
(a) GRSM-4QAM, (b) GRSM-16QAM, respectively.}
\label{fig:3}
\end{figure}

The transmitter has $N_t$ = 32 antennas and the receiver has $N_r$ = 8 antennas, with $N_t$ and single RF chains at the transmitter and receiver, respectively. 
PN is strong, with variance $\sigma_{\rm PN}^2 $= 0.1~rad \cite{b4,b5,b6}, and the AWGN variance is $\sigma^2 = E_s / (\text{SNR} \cdot \log_2(M))$.  
For GRSM-4QAM, GRSM-16QAM , symbol energy $E_s$ = 2, 10, respectively,\cite{b3}.
Fig.~\ref{fig:3a} and Fig.~\ref{fig:3b} illustrate the BER performance of different variants of GRSM-4QAM and GRSM-16QAM, respectively. Under CLO, the preserved received energy enables reliable spatial detection; however, the MQAM component is impaired, resulting in degraded overall BER performance (blue circles), with a flatter behavior in the denser 16QAM constellation. In contrast, under no-PN conditions, the BER of GRSM-MQAM is dominated by spatial errors, as concluded in \cite{b3}, which favors GRSM-16QAM over GRSM-4QAM (black dashed circles).
In Fig.~\ref{fig:3a}, the robust spatial detection leads to an accurate estimate of the decimal value $\hat{J}$ and correct symbol-pool selection in E-PN-GRSM-4QAM and E-PN-GRSM-16QAM (red triangles). Consequently, the $N_a$ spatial bits and the first $(m-1)$ bit of the MQAM sequence are reliably detected.
Moreover, the sparsity of the 4QAM symbol pools inherently mitigates CLO-induced PN, such that single-stage PN estimation provides only marginal additional gain, as indicated by the red squares in Fig.~\ref{fig:3a}. However, in Fig.~\ref{fig:3b} (red squares), PN introduces unequal relative distortion between the real and imaginary components in 16QAM constellation, which breaks the $\pi$-separated structure within PN-sensitive pools and increases intra-pool symbol errors due to the wrapping-function limitation in the one-stage compensation system.
For both E-PN-GRSM-4QAM and E-PN-GRSM-16QAM, (red dashed lines), the benchmark double-stage PN mitigation further suppresses the residual errors, showing the upper-bound performance achievable when Tx/Rx-PN are mitigated separately.

\section{CONCLUSION}
This work proposed (E-PN-GRSM-MQAM) exploiting the SM for PN mitigation through new mapping method benefiting the robust spatial detection and Hamming weights. Single- and double-stage symbol-assisted compensation were also considered for Tx/Rx-PN estimation. Results show that single-stage compensation improves PN robustness, while the benchmark double-stage scheme approaches near PN-free performance. Overall, the proposed approach offers a hardware-efficient PN mitigation solution for mmWave massive MIMO systems.

\end{document}